\documentclass[12pt]{article}
\usepackage{graphicx}
\usepackage{lscape}
\usepackage{amssymb}
\usepackage{appendix}
\usepackage{multirow}
\usepackage{citesort}

\begin{document}

\def\ds{\displaystyle}
\def\beq{\begin{equation}}
\def\eeq{\end{equation}}
\def\bea{\begin{eqnarray}}
\def\eea{\end{eqnarray}}
\def\beeq{\begin{eqnarray}}
\def\eeeq{\end{eqnarray}}
\def\ve{\vert}
\def\vel{\left|}
\def\ver{\right|}
\def\nnb{\nonumber}
\def\ga{\left(}
\def\dr{\right)}
\def\aga{\left\{}
\def\adr{\right\}}
\def\lla{\left<}
\def\rra{\right>}
\def\rar{\rightarrow}
\def\lrar{\leftrightarrow}
\def\nnb{\nonumber}
\def\la{\langle}
\def\ra{\rangle}
\def\ba{\begin{array}}
\def\ea{\end{array}}
\def\tr{\mbox{Tr}}
\def\ssp{{\Sigma^{*+}}}
\def\sso{{\Sigma^{*0}}}
\def\ssm{{\Sigma^{*-}}}
\def\xis0{{\Xi^{*0}}}
\def\xism{{\Xi^{*-}}}
\def\qs{\la \bar s s \ra}
\def\qu{\la \bar u u \ra}
\def\qd{\la \bar d d \ra}
\def\qq{\la \bar q q \ra}
\def\gGgG{\la g^2 G^2 \ra}
\def\q{\gamma_5 \not\!q}
\def\x{\gamma_5 \not\!x}
\def\g5{\gamma_5}
\def\sb{S_Q^{cf}}
\def\sd{S_d^{be}}
\def\su{S_u^{ad}}
\def\sbp{{S}_Q^{'cf}}
\def\sdp{{S}_d^{'be}}
\def\sup{{S}_u^{'ad}}
\def\ssp{{S}_s^{'??}}

\def\sig{\sigma_{\mu \nu} \gamma_5 p^\mu q^\nu}
\def\fo{f_0(\frac{s_0}{M^2})}
\def\ffi{f_1(\frac{s_0}{M^2})}
\def\fii{f_2(\frac{s_0}{M^2})}
\def\O{{\cal O}}
\def\sl{{\Sigma^0 \Lambda}}
\def\es{\!\!\! &=& \!\!\!}
\def\ap{\!\!\! &\approx& \!\!\!}
\def\ar{&+& \!\!\!}
\def\ek{&-& \!\!\!}
\def\kek{\!\!\!&-& \!\!\!}
\def\cp{&\times& \!\!\!}
\def\se{\!\!\! &\simeq& \!\!\!}
\def\eqv{&\equiv& \!\!\!}
\def\kpm{&\pm& \!\!\!}
\def\kmp{&\mp& \!\!\!}
\def\mcdot{\!\cdot\!}

% .........................................................

\def\simlt{\stackrel{<}{{}_\sim}}
\def\simgt{\stackrel{>}{{}_\sim}}

% .........................................................

\title{
         {\Large
                 {\bf
Nucleon tensor form factors induced by isovector and isoscalar
currents in QCD
                 }
         }
      }

\author{\vspace{1cm}\\
{\small T. M. Aliev$^1$ \thanks {e-mail:
taliev@metu.edu.tr}~\footnote{permanent address:Institute of
Physics,Baku,Azerbaijan}\,\,, K. Azizi$^2$ \thanks {e-mail:
kazizi@dogus.edu.tr}\,\,, M. Savc{\i}$^1$ \thanks
{e-mail: savci@metu.edu.tr}} \\
{\small $^1$ Physics Department, Middle East Technical University,
06531 Ankara, Turkey }\\
{\small$^2$ Department of Physics,  
Do\u gu\c s University, Ac{\i}badem-Kad{\i}k\"oy,  34722 Istanbul, Turkey}}

\date{}

\begin{titlepage}
\maketitle
\thispagestyle{empty}

\begin{abstract}
Using the most general form of the nucleon interpolating current, we calculate the tensor form factors of the nucleon within light cone QCD
sum rules. A comparison of our results on tensor form factors with those of  the chiral--soliton
model and lattice QCD is given.
\end{abstract}

%\vspace{1cm}
~~~PACS number(s): 11.55.Hx,  14.20.Dh
\end{titlepage}

\section{Introduction}

The main problem of QCD is to understand  the structure of hadrons and their properties
in terms of quarks and gluons. Nucleon charges defined as matrix elements of vector, axial and tensor currents between nucleon states contain complete information about quark 
structure of the nucleon. These charges are connected with the leading twist unpolarized $q(x)$, the helicity $\Delta q(x)$ and transversity $\delta q(x)$ 
parton  distribution functions (PDFs). The first two PDFs have been extensively investigated theoretically and experimentally in many works 
(for instance see \cite{R10601,R10602} and references therein as well as \cite{Pire1,Pire2,Pire3}). There is a big experimental  problem to measure the transversity of the nucleon because
 of its chiral odd nature.
Only, recently the tensor charge $\delta q(x)$ was extracted \cite{R10603} using the data from BELLE \cite{R10604}, HERMES
\cite{R10605} and COMPASS \cite{R10606} Collaborations. This extraction is based on analysis of the measured azimuthal asymmetries in semi-inclusive scattering and those 
in $e^+e^-\rar h_1h_2X$ processes. Since $\delta q(x)$ is a spin dependent PDF, it is interesting to investigate whether there is a "transversity crisis" similar
 to the case of "spin crisis" in $\Delta q(x)$. Therefore, reliable determination of nucleon tensor charge receives special attention.

Theoretically, tensor charges of hadrons are studied in different frameworks
such as, non--relativistic MIT bag model \cite{R10607}, SU(6) quark model
\cite{R10608}, quark model with axial vector dominance \cite{R10609},
lattice QCD \cite{R10610}, external field \cite{Han} and three point versions of QCD sum rules \cite{R10607}.  

In the present work, using the most general form of the 
nucleon interpolating field, we study the tensor form factors of nucleons within
light cone QCD sum rules (LCQSR). 
The LCQSR is based on the operator product expansion (OPE) over
twist of the operators near the light cone, while in the traditional QCD 
sum rules, the OPE is performed over dimensions of the operators. This
approach has been widely applied to  hadron physics (see for example
\cite{R10611}). 
Note that, the tensor form factors of nucleons up to $Q^2 \le
1~GeV^2$ (where $Q^2=-q^2$ is the Euclidean momentum transfer square) within
the $SU(3)$ chiral soliton model are studied in \cite{R10612} (see also \cite{Gamberg}). The anomalous
tensor form factors are studied within the same framework in \cite{R10613}.
These form factors are further studied in lattice QCD (see for
instance \cite{R10614}).  

The plan of this paper is as
follows. In section 2, we derive  sum rules for the tensor form factors of the nucleon
within LQCSR method. In section 3, we numerically analyze the sum rules for the tensor form
factors. A comparison of our results on form factors
with those existing in the literature is also presented in this section.

\section{Light cone sum rules for the nucleon tensor form factors}

This section is devoted to derivation of LCQSR for the nucleon tensor form
factors.  The matrix element of the 
tensor current between  initial and final nucleon states is parametrized in terms of
four form factors as follows \cite{R10601,R10614,R10615}:
\bea
\label{enucten01}
\lla N(p^\prime) \vel \bar{q} \sigma_{\mu\nu} q \ver N(p) \rra \es
\bar{u}(p^\prime) \Bigg\{
H_T (Q^2) i \sigma_{\mu\nu} - E_T(Q^2) {\gamma_\mu q_\nu - \gamma_\nu q_\mu
\over 2 m_N} + E_{1T}(Q^2) {\gamma_\mu {\cal P}_\nu - 
\gamma_\nu {\cal P}_\mu \over 2 m_N} \nnb \\
\ek \widetilde{H}_{T} (Q^2) {{\cal P}_\mu q_\nu - {\cal P}_\nu
q_\mu \over 2 m_N^2} \Bigg\}u(p)~,
\eea
where $q_\mu = (p-p^\prime)_\mu$, ${\cal P}_\mu = (p+p^\prime)_\mu$, and 
$q^2=-Q^2$. From T--invariance it follows  that $ E_{1T}(Q^2)=0$.

In order to calculate the remaining three tensor form factors within LCQSR, 
we consider the  correlation function,
\bea
\label{enucten02}
\Pi_{\mu\nu} (p,q) = i \int d^4x e^{iqx} \lla 0 \vel T\{J^N(0) J_{\mu\nu} (x)
\} \ver N(p) \rra~.
\eea
This correlation function describes  transition of the initial nucleon to
the final nucleon with the help of the tensor current. The most general form
of the nucleon interpolating field is given as,
\bea
\label{enucten03}
J^N(x) = 2 \varepsilon^{abc} \sum_{i=1}^2 \Big[q^{Ta}(x) C A_1^i
q^{'b}(x) \Big] A_2^i q^c (x)~,
\eea 
where $C$ is the charge conjugation operator,
$A_1^1=I$, $A_1^2=A_2^1=\gamma_5$, $A_2^2=t$ with $t$ being
an arbitrary parameter and $t=-1$ corresponds to the Ioffe current
and $a,b,c$ are the color indices. The quark
flavors are $q=u$, $q^{'}=d$ for the proton and $q=d$, $q^{'}=u$ for the
neutron. The tensor current is chosen as,
\bea
\label{enucten04}
J_{\mu\nu} = \bar{u} \sigma_{\mu\nu} u \pm \bar{d} \sigma_{\mu\nu} d~,
\eea
where the upper and lower signs correspond to the isosinglet and  isovector
cases, respectively.

In order to obtain sum rules for the form factors, it is necessary to
calculate the correlation function in terms of quarks and gluons  on  one
side (QCD side), and in terms of hadrons  on the other side (phenomenological side).
These two representations of the correlation function are then equated. The
final step in this method is to apply the Borel transformation, which is needed to
suppress the higher states  and the continuum contributions. 

Following this strategy, we start to calculate the phenomenological part.
Saturating the correlation function with a full set of hadrons carrying
the same quantum numbers as nucleon and isolating the contributions of the
ground state, we get
\bea
\label{enucten05}  
\Pi_{\mu\nu} (p,q) = 
{\lla 0 \vel J^N(0) \ver N(p^\prime) \rra
\lla N(p^\prime) \vel J_{\mu\nu} \ver N(p) \rra\over
m_N^2-p^{\prime 2}} + \cdots~,
\eea
where dots stands for contributions of higher states and continuum. The matrix element $\lla 0
\vel J^N(0) \ver N(p^\prime) \rra$ entering  Eq. (\ref{enucten05}) is
defined as
\bea
\label{enucten06}
\lla 0 \vel J^N(0) \ver N(p^\prime) \rra = \lambda_N u(p)~,
\eea
where $\lambda_N$ is the residue of the nucleon. Using Eqs.
(\ref{enucten01}), (\ref{enucten05}) and (\ref{enucten06}), and performing
summation over spins of the nucleon, we get,
\bea
\label{enucten07}
\Pi_{\mu\nu} \es {\lambda_N \over m_N^2-p^{\prime 2}} ({\rlap/{p}}^{_{'}} + m_N)
\Bigg\{ H_T (Q^2) i \sigma_{\mu\nu} - E_T (Q^2) {\gamma_\mu q_\nu -
\gamma_\nu q_\mu \over 2 m_N} \nnb \\
\ek \widetilde{H}_{T} (Q^2) {{\cal P}_\mu q_\nu - 
{\cal P}_\nu q_\mu \over 2 m_N^2} \Bigg\} u(p)~.
\eea
 From Eq. (\ref{enucten07}) we see that there are many structures, and all of
them play equal role for determination of the tensor form factors of the
nucleon. In practical applications, it is more useful to work with
$\widetilde{E}_{T}(Q^2) = E_T(Q^2) + 2 \widetilde{H}_{T}(Q^2)$ rather than
$E_T(Q^2)$. For this reason, we choose the structures $\sigma_{\mu\nu}$,
$p_\mu q_\nu$ and $p_\mu q_\nu \rlap/{q}$ for obtaining the
sum rules for the form factors $H_T$, $\widetilde{E}_T$ and $\widetilde{H}_{T}$,
respectively.

The correlation function $\Pi_{\mu\nu} (p,q)$ is also calculated in terms of
quarks and gluons  in deep Eucledian domain $p^{'2} = (p-q)^2
<< 0$. After simple calculations, we get the following expression for the correlation function for proton case:
\bea
\label{enucten08}
\ga \Pi_{\mu\nu} \dr_\rho \es {i \over2} \int d^4x e^{iqx} \sum_{i=1}^2 \Bigg\{
\ga C A_1^i\dr_{\alpha\tau} \Big[ A_2^i S_u(-x) \sigma_{\mu\nu} 
\Big]_{\rho\sigma}
4 \epsilon^{abc} \lla 0 \vel u_\alpha^a(0) u_\sigma^b(x)
d_\tau^c(0) \ver N (p)\rra \nnb \\ 
\ar \ga A_2^i \dr_{\rho\alpha} \Big[ \ga C A_1^i\dr^T S_u(-x)
\sigma_{\mu\nu}\Big]_{\tau\sigma} 4 \epsilon^{abc} \lla 0 \vel
u_\alpha^a(x)
u_\sigma^b(x) d_\tau^c(0) \ver N (p)\rra \nnb \\
\kpm \ga A_2^i \dr_{\rho\sigma} \Big[ C A_1^i S_d(-x)
\sigma_{\mu\nu}\Big]_{\alpha\tau} 4 \epsilon^{abc} \lla 0 \vel u_\alpha^a(0)
u_\sigma^b(0)  d_\tau^c(x) \ver N (p)\rra \Bigg\}~.
\eea
Obviously, the correlation function for the neutron case can easily be
obtained by making the replacement $u \lrar d$.

From Eq. (\ref{enucten08}) it is clear that in order to calculate  the correlation
function from QCD side, we need to know the matrix element,
\bea
\label{nolabel}
4 \varepsilon^{abc} \lla 0 \vel u_\alpha^a(a_1 x) u_\sigma^b(a_2 x)
d_\tau^a(a_3 x) \ver N(p) \rra~,\nnb
\eea
where $a_1$, $a_2$ and $a_3$ determine the fraction of the nucleon
momentum carried by the corresponding quarks. This matrix element is the
main nonperturbative ingredient of the sum rules and it is defined in terms of the
nucleon distribution amplitudes (DAs). The nucleon DAs are studied in detail in
\cite{R10616,R10617,R10618}.

The light cone expanded expression for the light quark propagator $S_q(x)$   
is given as,
\bea
\label{enucten09}
S_q(x) = \frac{i \rlap/x}{2 \pi^2 x^4} - \frac{\la q \bar{q} \ra}{12} \Bigg(
1+\frac{m_0^2 x^2}{16} \Bigg)
- i g_s \int_0^1 dv \Bigg[ \frac{\rlap/x}{16 \pi^2 x^4} G_{\mu\nu}
\sigma^{\mu\nu} - v x^\mu G_{\mu\nu} \gamma^\nu \frac{i}{4 \pi^2 x^2}
\Bigg]~,
\eea
where the mass of the light quarks are neglected, $m_0^2=(0.8 \pm
0.2)~GeV^2$ \cite{R10619} and $G_{\mu\nu}$ is the gluon field strength
tensor. The terms containing $G_{\mu\nu}$ give contributions to four-- and
five--particle distribution functions. These contributions are negligibly
small (for more detail see \cite{R10616,R10617,R10618}), and therefore in
further analysis, we will neglect these terms. Moreover, Borel
transformation kills the terms proportional to the quark condensate, and as
a result only the first term is relevant for our discussion.

Using the explicit expressions of DAs for the proton and light quark
propagators, performing Fourier transformation and then applying 
Borel transformation with respect to the variable $p^{'2}=(p-q)^2$, which
suppresses the contributions of continuum and higher states, and choosing
the coefficients of the structures $\sigma_{\mu\nu}$, $p_\mu q_\nu$
and $p_\mu q_\nu\rlap/{q}$, we get the following sum rules for the
tensor form factors of nucleon:\\\\

\bea
\label{enucten10}
H_T (Q^2) \es {1 \over 2 m_N \lambda_N} e^{m_N^2/M^2} \Bigg\{
%(....)^2 -------------------------------------------
\int_{x_0}^1 {dt_2\over t_2} e^{-s(t_2)/M^2} \Big[ (1-t) 
F_{H_T}^{1}(t_2) + (1+t) F_{H_T}^{2}(t_2) \Big] \nnb \\
\kpm \int_{x_0}^1 {dt_3\over t_3} e^{-s(t_3)/M^2} \Big[ (1-t) 
F_{H_T}^{3}(t_3) + (1+t) F_{H_T}^{4}(t_3) \Big] \nnb \\
%----------------------------------------------------
%(....)^2 Px ----------------------------------------
\ar \int_{x_0}^1 {dt_2\over t_2} e^{-s(t_2)/M^2} \Big[ (1-t) 
F_{H_T}^{5}(t_2) + (1+t) F_{H_T}^{6}(t_2) \Big] \nnb \\
\kpm \int_{x_0}^1 {dt_3\over t_3} e^{-s(t_3)/M^2} \Big[ (1-t) 
F_{H_T}^{7}(t_3) + (1+t) F_{H_T}^{8}(t_3) \Big] \nnb \\
%----------------------------------------------------
%(....)^2 Px^2 ----------------------------------------
\ar \int_{x_0}^1 {dt_2\over t_2} e^{-s(t_2)/M^2} (1-t)
F_{H_T}^{9}(t_2) \nnb \\
\kpm \int_{x_0}^1 {dt_3\over t_3} e^{-s(t_3)/M^2} 
(1+t) F_{H_T}^{10}(t_3) \nnb \\
%----------------------------------------------------
%(....)^4 -------------------------------------------
\ar {1 \over M^2} \int_{x_0}^1 {dt_2\over t_2^2}
e^{-s(t_2)/M^2} \Big[ (1-t) F_{H_T}^{11}(t_2) +
(1+t) F_{H_T}^{12}(t_2) \Big] \nnb \\
\ar {1 \over Q^2+x_0^2 m_N^2} e^{-s_0/M^2}
\Big[ (1-t) F_{H_T}^{11}(x_0) + (1+t) F_{H_T}^{12}(x_0) \Big] \nnb \\
\kpm {1 \over M^2} \int_{x_0}^1 {dt_3\over t_3^2}
e^{-s(t_3)/M^2} \Big[ (1-t) F_{H_T}^{13}(t_3) +
(1+t) F_{H_T}^{14}(t_3) \Big] \nnb \\
\kpm {1 \over Q^2+x_0^2 m_N^2} e^{-s_0/M^2}
\Big[ (1-t) F_{H_T}^{13}(x_0) + (1+t) F_{H_T}^{14}(x_0) \Big] \nnb \\
%----------------------------------------------------
%(....)^4 (Px)^2 ------------------------------------
\ar {1 \over M^2} \int_{x_0}^1 {dt_2\over t_2^2}
e^{-s(t_2)/M^2} \Big[ (1-t) F_{H_T}^{15}(t_2) +
(1+t) F_{H_T}^{16}(t_2) \Big] \nnb \\
\ar {1 \over Q^2+x_0^2 m_N^2} e^{-s_0/M^2}
\Big[ (1-t) F_{H_T}^{15}(x_0) + (1+t) F_{H_T}^{16}(x_0) \Big] \nnb \\
\kpm {1 \over M^2} \int_{x_0}^1 {dt_3\over t_3^2}
e^{-s(t_3)/M^2} (1+t) F_{H_T}^{17}(t_3) \nnb \\
\kpm {1 \over Q^2+x_0^2 m_N^2} e^{-s_0/M^2}
(1+t) F_{H_T}^{17}(x_0) \Bigg\}~,
%----------------------------------------------------
\eea
where
\bea
\label{??}
F_{H_T}^{1}(t_2) \es  \int_0^{1-t_2} dt_1 \Bigg\{ 
{2 m_N^2 \over t_2} \Big[\widetilde{\cal T}_1^M + t_2^2 (\widetilde{\cal
P}_1 - 3 \widetilde{\cal T}_3 - \widetilde{\cal T}_4)\Big]
(t_1,t_2,1-t_1-t_2) \nnb \\ 
\ar {2(Q^2+ m_N^2 t_2^2) \over t_2} 
\widetilde{\cal T}_1 (t_1,t_2,1-t_1-t_2)\Bigg\}~, \nnb \\
F_{H_T}^{2}(t_2) \es % -  
\int_0^{1-t_2} dt_1 \Bigg\{
{m_N^2 \over t_2} \Big[\widetilde{\cal V}_1^M - \widetilde{\cal A}_1^M - t_2^2
(\widetilde{\cal A}_2 + 3 \widetilde{\cal A}_3 + \widetilde{\cal V}_2 +
3 \widetilde{\cal V}_3)\Big](t_1,t_2,1-t_1-t_2) \nnb \\
\ek {Q^2+ m_N^2 t_2^2 \over t_2} \Big[\widetilde{\cal A}_1 - \widetilde{\cal
V}_1 \Big](t_1,t_2,1-t_1-t_2)\Bigg\}~, \nnb \\
F_{H_T}^{3}(t_3) \es %\pm  
\int_0^{1-t_3} dt_1 
{1\over t_3} \Big[ m_N^2 (\widetilde{\cal A}_1^M + \widetilde{\cal V}_1^M) +
m_N^2 t_3^2 (\widetilde{\cal A}_3 - \widetilde{\cal V}_3) +
Q^2 (\widetilde{\cal A}_1 + \widetilde{\cal V}_1)\Big] (t_1,1-t_1-t_3,t_3)~, \nnb \\
F_{H_T}^{4}(t_3) \es %\pm
\int_0^{1-t_3} dt_1
\Big[ {2\over t_3} ( m_N^2 \widetilde{\cal T}_1^M + Q^2 \widetilde{\cal T}_1)
- {m_N^2 t_3 \over 2} (2 \widetilde{\cal P}_1 - 2 \widetilde{\cal S}_1 + 2
\widetilde{\cal T}_1 - \widetilde{\cal T}_2 - \widetilde{\cal T}_4 ) \Big]
(t_1,1-t_1-t_3,t_3)~, \nnb \\
F_{H_T}^{5}(t_2) \es % -
{m_N^2\over 2} 
\int_1^{t_2} d\rho \int_0^{1-\rho} dt_1
\Big[ 4 \widetilde{\cal T}_4 + 4 \widetilde{\cal T}_5
- 3 \widetilde{\cal T}_6 + 12 \widetilde{\cal T}_7 -  4 \widetilde{\cal S}_2
\Big](t_1,\rho,1-t_1-\rho)~, \nnb \\ 
F_{H_T}^{6}(t_2) \es m_N^2 
\int_1^{t_2} d\rho \int_0^{1-\rho} dt_1
\Big[2 \widetilde{\cal A}_2 - \widetilde{\cal A}_4 + 2 \widetilde{\cal A}_5
+ 2 \widetilde{\cal V}_2 + \widetilde{\cal V}_4 -2 \widetilde{\cal N}_5
\Big](t_1,\rho,1-t_1-\rho)~, \nnb \\
F_{H_T}^{7}(t_3) \es %\pm
{m_N^2 \over 2} 
\int_1^{t_3} d\rho \int_0^{1-\rho} dt_1
\Big[2 \widetilde{\cal A}_2 - \widetilde{\cal A}_4 + \widetilde{\cal A}_5
- 2 \widetilde{\cal V}_2 - \widetilde{\cal V}_4 + \widetilde{\cal V}_5
\Big](t_1,1-t_1-\rho,\rho)~, \nnb \\
F_{H_T}^{8}(t_3) \es %\mp
{m_N^2 \over 2} 
\int_1^{t_3} d\rho \int_0^{1-\rho} dt_1
\Big[2 \widetilde{\cal T}_2 + 2 \widetilde{\cal T}_5 - \widetilde{\cal T}_6
- 2 \widetilde{\cal P}_2 - 2 \widetilde{\cal S}_4
\Big](t_1,1-t_1-\rho,\rho)~, \nnb \\
F_{H_T}^{9}(t_2) \es {m_N^2 \over 2}  \int_1^{t_2} d\lambda
\int_1^{\lambda} d\rho \int_0^{1-\rho} dt_1 {1\over \rho} \widetilde{\cal
T}_6 (t_1,\rho,1-t_1-\rho)~, \nnb \\
F_{H_T}^{10}(t_3) \es %\mp
m_N^2  \int_1^{t_3} d\lambda
\int_1^{\lambda} d\rho \int_0^{1-\rho} dt_1
{1\over \rho} \widetilde{\cal 
T}_6 (t_1,1-t_1-\rho,\rho)~, \nnb \\
F_{H_T}^{11}(t_2) \es 2 m_N^2  \int_0^{1-t_2} dt_1 
{Q^2+m_N^2 t_2^2\over t_2} 
\widetilde{\cal T}_1^M (t_1,t_2,1-t_1-t_2)~, \nnb \\
F_{H_T}^{12}(t_2) \es % -
m_N^2  \int_0^{1-t_2} dt_1 
{Q^2+m_N^2 t_2^2\over t_2} \Big[ \widetilde{\cal V}_1^M - 
\widetilde{\cal A}_1^M \Big](t_1,t_2,1-t_1-t_2)~, \nnb \\
F_{H_T}^{13}(t_3) \es %\pm
m_N^2  \int_0^{1-t_3} dt_1 
{Q^2 \over t_3} 
\Big[\widetilde{\cal A}_1^M + \widetilde{\cal V}_1^M\Big]
(t_1,1-t_1-t_3,t_3)~, \nnb \\
F_{H_T}^{14}(t_3) \es %\pm
m_N^2   \int_0^{1-t_3} dt_1 
{2 Q^2-m_N^2 t_3^2\over t_3} \widetilde{\cal T}_1^M (t_1,1-t_1-t_3,t_3)~, \nnb \\ 
F_{H_T}^{15}(t_2) \es {m_N^2 \over 2}  \int_1^{t_2} d\lambda 
\int_1^{\lambda} d\rho \int_0^{1-\rho}
dt_1 {1\over \rho}\Big[(Q^2+ m_N^2 \rho^2) \widetilde{\cal T}_6 + 
8  m_N^2 \rho^2 \widetilde{\cal T}_8 \Big](t_1,\rho,1-t_1-\rho)~, \nnb \\
\label{??}
F_{H_T}^{16}(t_2) \es 2 m_N^4 (1 +t) \int_1^{t_2} d\lambda 
\int_1^{\lambda} d\rho \int_0^{1-\rho} dt_1
\rho \Big[\widetilde{\cal A}_6 + \widetilde{\cal
V}_6\Big](t_1,\rho,1-t_1-\rho)~, \nnb \\
F_{H_T}^{17}(t_3) \es %\mp
m_N^2 
\int_1^{t_3} d\lambda \int_1^{\lambda} d\rho \int_0^{1-\rho} dt_1 
{1\over \rho} \Big[ m_N^2 \rho^2 \widetilde{\cal T}_8  - 
Q^2 \widetilde{\cal T}_6\Big](t_1,1-t_1-\rho,\rho)~.
%
% F_{H_T}^{1242} \es 0
%
\eea
%-----------------------------------

For the form factor $\widetilde{E}_{T} (Q^2)$ we obtain the following sum
rule:  

\bea
\label{enucten11}
\widetilde{E}_{T} (Q^2) \es {1 \over m_N \lambda_N} e^{m_N^2/M^2} \Bigg\{
%(....)^4 (Px)^2 -------------------------------------------
{1 \over M^2} \int_{x_0}^1 {dt_2\over t_2^2}
e^{-s(t_2)/M^2} (1-t) F_{\widetilde{E}_{T}}^1 (t_2) \nnb \\
\ar {1 \over Q^2+x_0^2 m_N^2} e^{-s_0/M^2}
(1-t) F_{\widetilde{E}_{T}}^1 (x_0) \nnb \\
%----------------------------------------------------
\ar {1 \over M^2} \int_{x_0}^1 {dt_2\over t_2^2}
e^{-s(t_2)/M^2} (1-t) F_{\widetilde{E}_{T}}^3(t_2) \nnb \\
\ar {1 \over Q^2+x_0^2 m_N^2} e^{-s_0/M^2}
(1-t) F_{\widetilde{E}_{T}}^3(x_0) \nnb \\
\kpm {1 \over M^2} \int_{x_0}^1 {dt_3\over t_3^2}
e^{-s(t_3)/M^2} (1-t) F_{\widetilde{E}_{T}}^4(t_3) \nnb \\
\kpm {1 \over Q^2+x_0^2 m_N^2} e^{-s_0/M^2}
(1-t) F_{\widetilde{E}_{T}}^4(x_0) \nnb \\
%(....)^4 -------------------------------------------
\ar {1 \over M^2} \int_{x_0}^1 {dt_2\over t_2^2}
e^{-s(t_2)/M^2} (1-t) F_{\widetilde{E}_{T}}^5 (t_2) \nnb \\
\ar {1 \over Q^2+x_0^2 m_N^2} e^{-s_0/M^2}
(1-t) F_{\widetilde{E}_{T}}^5(x_0) \nnb \\
\kpm {1 \over M^2} \int_{x_0}^1 {dt_3\over t_3^2}
e^{-s(t_3)/M^2} (1-t) F_{\widetilde{E}_{T}}^6(t_3) \nnb \\
\kpm {1 \over Q^2+x_0^2 m_N^2} e^{-s_0/M^2}
(1-t) F_{\widetilde{E}_{T}}^6(x_0) \nnb \\
%----------------------------------------------------
%(....)^2 -------------------------------------------
\ar \int_{x_0}^1 {dt_2\over t_2} e^{-s(t_2)/M^2} (1-t)
F_{\widetilde{E}_{T}}^7 (t_2) \nnb \\
\kpm \int_{x_0}^1 {dt_3\over t_3} e^{-s(t_3)/M^2} (1-t)
F_{\widetilde{E}_{T}}^8 (t_3) \Bigg\}~,
%----------------------------------------------------
\eea
where
\bea
\label{??}
F_{\widetilde{E}_{T}}^1(t_2) \es - 4 m_N^2 
\int_1^{t_2} d\lambda
\int_1^{\lambda} d\rho \int_0^{1-\rho} dt_1
\widetilde{\cal T}_6(t_1,\rho,1-t_1-\rho)~, \nnb \\
F_{\widetilde{E}_{T}}^3(t_2) \es - 4 m_N^2 
\int_1^{t_2} d\rho \int_0^{1-\rho} dt_1 \rho \Big[\widetilde{\cal T}_2 +
\widetilde{\cal T}_4 \Big](t_1,\rho,1-t_1-\rho)~, \nnb \\
F_{\widetilde{E}_{T}}^4(t_3) \es % \pm
4 m_N^2 
\int_1^{t_2} d\rho \int_0^{1-\rho} dt_1 \rho \Big[\widetilde{\cal A}_2 -
\widetilde{\cal V}_2 \Big](t_1,1-t_1-\rho,\rho)~, \nnb \\
F_{\widetilde{E}_{T}}^5 (t_2) \es 8  m_N^2  \int_0^{1-t_2} dt_1
\widetilde{\cal T}_1^M (t_1,t_2,1-t_1-t_2)~, \nnb \\
F_{\widetilde{E}_{T}}^6 (t_3) \es % \pm
4 m_N^2  \int_0^{1-t_3} dt_1 \Big[\widetilde{\cal A}_1^M +
\widetilde{\cal V}_1^M
\Big](t_1,1-t_1-t_3,t_3)~, \nnb \\
F_{\widetilde{E}_{T}}^7 (t_2) \es 8   \int_0^{1-t_2} dt_1 
\widetilde{\cal T}_1 (t_1,t_2,1-t_1-t_2))~, \nnb \\
F_{\widetilde{E}_{T}}^8 (t_3) \es % \pm
4  \int_0^{1-t_3} dt_1
\Big[\widetilde{\cal A}_1 +
\widetilde{\cal V}_1 \Big](t_1,1-t_1-t_3,t_3)~. \nnb
\eea
%---------------------------

Finally for the form factor $\widetilde{H}_{T}(Q^2)$ we get the following
sum rule:

\bea
\label{enucten12}
\widetilde{H}_{T}(Q^2) \es {1 \over m_N^2 \lambda_N} e^{m_N^2/M^2} \Bigg\{
%(....)^4 (Px) -------------------------------------------
{1\over M^2} \int_{x_0}^1 {dt_2\over t_2^2}
e^{-s(t_2)/M^2} (1-t) F_{\widetilde{H}_{T}}^1(t_2) 
+ {1 \over Q^2+x_0^2 m_N^2} e^{-s_0/M^2}
(1-t) F_{\widetilde{H}_{T}}^1(x_0) \nnb \\
\kpm {1 \over M^2} \int_{x_0}^1 {dt_3\over t_3^2}
e^{-s(t_3)/M^2} (1-t) F_{\widetilde{H}_{T}}^2(t_3) 
\pm {1 \over Q^2+x_0^2 m_N^2} e^{-s_0/M^2}
(1-t) F_{\widetilde{H}_{T}}^2(x_0)\Bigg\}~,
\eea
where
\bea
\label{??}
F_{\widetilde{H}_{T}}^1(t_2) \es 4 m_N \int_1^{t_2} d\rho \int_0^{1-\rho} dt_1
\Big[\widetilde{\cal T}_2 + \widetilde{\cal T}_4
\Big](t_1,\rho,1-t_1-\rho)~, \nnb \\
F_{\widetilde{H}_{T}}^2 (t_3)\es 4 m_N \int_1^{\rho} d\rho \int_0^{1-\rho} dt_1
\Big[- \widetilde{\cal A}_2 + \widetilde{\cal V}_2
\Big](t_1,1-t_1-\rho,\rho)~, \nnb
\eea
%---------------------------------
and we use

\bea
\widetilde{\cal V}_2(t_i) \es V_1(t_i) - V_2(t_i) - V_3(t_i)~, \nnb \\
\widetilde{\cal A}_2(t_i) \es - A_1(t_i) + A_2(t_i) - A_3(t_i)~, \nnb \\
\widetilde{\cal A}_4(t_i) \es - 2 A_1(t_i) - A_3(t_i) - A_4(t_i)
+ 2 A_5(t_i)~, \nnb \\
\widetilde{\cal A}_5(t_i) \es A_3(t_i) - A_4(t_i)~, \nnb \\ 
\widetilde{\cal A}_6(t_i) \es A_1(t_i) - A_2(t_i) + A_3(t_i) +
A_4(t_i) - A_5(t_i) + A_6(t_i)~, \nnb \\
\widetilde{\cal T}_2(t_i) \es T_1(t_i) + T_2(t_i) - 2 T_3(t_i)~, \nnb \\
\widetilde{\cal T}_4(t_i) \es T_1(t_i) - T_2(t_i) - 2 T_7(t_i)~, \nnb \\
\widetilde{\cal T}_5(t_i) \es - T_1(t_i) + T_5(t_i) + 2 T_8(t_i)~, \nnb \\
\widetilde{\cal T}_6(t_i) \es 2 \Big[T_2(t_i) - T_3(t_i) - T_4(t_i) +
T_5(t_i) + T_7(t_i) + T_8(t_i)\Big]~, \nnb \\
\widetilde{\cal T}_7(t_i) \es T_7(t_i) - T_8(t_i)~, \nnb \\
\widetilde{\cal S}_2(t_i) \es S_1(t_i) - S_2(t_i)~, \nnb \\ 
\widetilde{\cal P}_2(t_i) \es P_2(t_i) - P_1(t_i)~, \nnb
\eea
In these expressions, we also  use
\bea
\label{nolabel}
{\cal F}(x_i) \es {\cal F}(x_1,x_2,1-x_1-x_2) ~, \nnb \\
{\cal F}(x_i^{'}) \es {\cal F}(x_1,1-x_1-x_3,x_3) ~, \nnb \\ 
s(x,Q^2) \es (1-x) m_N^2 + {(1-x)\over x} Q^2~, \nnb
\eea
where $x_0(s_0,Q^2)$ is the solution to the equation $s(x_0,Q^2) = s_0$.

The residue $\lambda_N$ is determined from two--point sum rule. For the
general form of the interpolating current, it is calculated in \cite{R10620},
whose expression is given as
\bea
\label{enucten13}
\lambda_N^2 \es e^{m_N^2/M^2} \Bigg\{\frac{M^6}{256 \pi^4} E_2(x) (5+2 t +
t^2) - \frac{\la \bar{u}u \ra}{6} \Big[6 (1-t^2) \la \bar{d}d \ra -
(1-t)^2 \la \bar{u}u \ra \Big] \nnb \\
\ar \frac{m_0^2}{24 M^2} \la \bar{u}u \ra \Big[12 (1-t^2) \la \bar{d}d
\ra - (1-t)^2 \la \bar{u}u \ra \Big]\Bigg\}~, \nnb
\eea
where
\bea
E_2(s_0/M^2) = 1-e^{s_0/M^2} \sum_{i=0}^2 \frac{(s_0/M^2)^i}{i!}~. \nnb
\eea
The Borel transformations are implemented by the following subtraction rules
\cite{R10616,R10617,R10618},
\bea
\label{e8315}
\int dx {\rho(x) \over (q-xp)^2} &\rar& -
\int {dx \over x}\rho(x) e^{-s(x)/M^2}~, \nnb \\
\int dx {\rho(x)\over (q-xp)^4} &\rar& {1\over M^2}
\int {dx\over x^2}\rho(x) e^{-s(x)/M^2}
+ {\rho(x_0)\over Q^2+x_0^2 m_N^2}e^{-s_0/M^2}~,\nnb \\
\int dx {\rho(x)\over (q-xp)^6} &\rar& - {1 \over 2 M^2}
\int {dx\over x^3}\rho(x) e^{-s(x)/M^2} - {1\over 2} {\rho(x_0) \over x_0 (Q^2
+ x_0^2 m_N^2) M^2} e^{-s_0/M^2} \nnb \\
&\phantom{\rar}& + {1\over 2} {x_0^2 \over Q^2 + x_0^2 m_N^2} \Bigg\{ 
{d\over dx_0} \Bigg[{1\over x_0} {\rho(x_0)\over Q^2+x_0^2 m_N^2}\Bigg]\Bigg\} e^{-s_0/M^2}~.
\eea

\section{Numerical analysis of the sum rules for the tensor form factors of
nucleon}

In this section, numerical results of the tensor form factors of nucleon are
presented. It follows from sum rules for the form factors that the main
input parameters are the DAs of nucleon, whose explicit expressions and the
values of the parameters $f_N$, $\lambda_1$, $\lambda_2$, $f_1^u$, $f_1^d$,
$A_1^u$, and $V_1^d$ in the DAs are all given in \cite{R10616,R10617,R10618}.

In the numerical analysis, we use two different sets of parameters:

a) All eight nonperturbative parameters $f_N$, $\lambda_1$,
$\lambda_2$, $f_1^u$, $f_1^d$, $f_2^d$, $A_1^u$ and $V_1^d$ are estimated
from QCD sum rules (set 1).

b) Requiring that all higher conformal spin contributions vanish, fixes
five $A_1^u$, $V_1^d$, $f_1^u$, $f_1^d$, and $A_2^d$, and the values of the
parameters $f_N$, $\lambda_1$, $\lambda_2$ are taken from QCD sum rules. This set is called asymptotic set or set2.

The values of all eight nonperturbative parameters (see for example
\cite{R10621}) are  presented in Table 1,

\begin{table}[tbh]
\renewcommand{\arraystretch}{1.3}
\addtolength{\arraycolsep}{-0.5pt}
\small
$$
\begin{array}{|l|c|c|}
\hline \hline
           &  \mbox{Set 1}                       &  \mbox{Asymptotic set (set2)}               \\  \hline \hline
 f_N       & ( 5.0 \pm 0.5)\times 10^{-3}~GeV^2  &  ( 5.0 \pm 0.5)\times 10^{-3}~GeV^2                    \\
 \lambda_1 & (-2.7 \pm 0.9)\times 10^{-2}~GeV^2  &   (-2.7 \pm 0.9)\times 10^{-2}~GeV^2                   \\
 \lambda_2 & ( 5.4 \pm 1.9)\times 10^{-2}~GeV^2  & ( 5.4 \pm 1.9)\times 10^{-2}~GeV^2                    \\
 A_1^u     & 0.38 \pm 0.15                       & 0                         \\
 V_1^d     & 0.23 \pm 0.03                       & 1/3                          \\
 f_1^d     & 0.40 \pm 0.05                       & 1/3                                            \\
 f_2^d     & 0.22 \pm 0.05                       & 4/15                                           \\
 f_1^u     & 0.07 \pm 0.05                       & 1/10                                          \\ \hline \hline
\end{array}
$$
\caption{The values of eight input parameters entering  the DAs of
nucleon.}
\renewcommand{\arraystretch}{1}
\addtolength{\arraycolsep}{-1.0pt}
\end{table}
 
The next input parameter of the LCQSR for the tensor form factors is the
continuum threshold $s_0$. This parameter is determined from the two--point sum rules
whose  value is in the domain $s_0=(2.25-2.50)~GeV^2$. The sum rules
contain also two extra auxiliary parameters, namely Borel parameter $M^2$ and
the parameter $t$ entering  the expression of the interpolating current
for nucleon. Obviously, any physical quantity should be independent of these
artificial parameters. Therefore, we try to find such regions of $M^2$ and
$t$, where the tensor form factors are insensitive to the variation of these
parameters.

Firstly, we try to obtain the working region of $M^2$, where the tensor form
factors are independent of it, at fixed values of $s_0$ and $t$. As an
example, in Figs. (1) and (2) we present  the dependence of the tensor form 
factor $H_T(Q^2)$ induced by the isoscalar current on $M^2$ at different fixed
values of $Q^2$ and $t$, and at
$s_0=2.25~GeV^2$ and $s_0=2.50~GeV^2$ for sets 1 and 2, respectively. From these figures, we see
 that $H_T(Q^2)$ is practically independent of $M^2$ at
fixed values of the parameters $Q^2$, $s_0$ and $t$ for both sets 1 and 2. Our calculations also show that the results are approximately the same for two sets, 
therefore in further discussion, we present the results only for set 1.
We  perform similar analysis also at $s_0=2.40~GeV^2$ and observe that the
results change maximally about 5\%. The upper limit of $M^2$ is determined
by requiring that the series of light cone expansion with increasing twist
converges, i.e., higher twist contributions should be small.
Our analysis indeed confirms that the twist--4 contributions to the sum rules
constitute maximally about 8\% of the total result when $M^2 \le
2.5~GeV^2$. The lower bound of $M^2$ is determined by requiring that the
contribution of the highest power of $M^2$ is less than, say, 30\% of the
higher powers of $M^2$. Our numerical analysis shows that this condition is
satisfied when $M^2\ge 1.0~GeV^2$. Hence, the working region of $M^2$ is
decided to be  in the interval $1.0~GeV^2 \le M^2 \le 2.5~GeV^2$. The working region
of the parameter $t$ is determined in such a way that the tensor form factors
are also independent of it. Our numerical analysis shows that the form factors
are insensitive to $\cos\theta$ (with $t=\tan\theta$) when it varies in the region 
$-0.5 \le \cos\theta \le 0.3$.

In Figs. (3)--(5) we present the dependence of the form factors $H_T(Q^2)$,
$\widetilde{E}_{T}(Q^2)$ and $\widetilde{H}_{T}(Q^2)$ on $Q^2$ at $s_0=2.25~GeV^2$,
$M^2=1.2~GeV^2$ and fixed values of $t$, respectively, using the central
values of all input parameters in set 1 for the isoscalar current.
For a comparison, we also present  the predictions of self consistent chiral soliton model 
\cite{R10612} and lattice QCD calculations \cite{R10614,R10615} in
these figures (note that,
chiral soliton model result exists only for $H_T(Q^2)$).

We see from Fig. (3) that, our results on  $H_T(Q^2)$ are close to the
lattice QCD results for $Q^2 \ge 2.0~GeV^2$, while the results of two models  differ
from each other in the region $1.0~GeV^2 \le Q^2 \le 2.0~GeV^2$. 
Our  and  lattice QCD  results differ considerably from the
predictions of the chiral soliton model. It also follows from these figures that the form factors get positive (negative)
at negative (positive) values of the parameter $t$.

In Figs. (6), (7) and (8), we present the dependence of the form factors
$H_T(Q^2)$ , $\widetilde{E}_{T}(Q^2)$ and $\widetilde{H}_{T}(Q^2)$ for the isovector current, i.e., for
the $\bar{u} \sigma_{\mu\nu} u - \bar{d} \sigma_{\mu\nu} d$ current. Our
observations for set 1 can be summarized as follows:

\begin{itemize}

\item The $Q^2$ dependence of $H_T(Q^2)$ is similar to the isoscalar current case,
but the values are slightly larger compared to the previous case.

\item Similar to the isoscalar case, the form factors $H_T(Q^2)$ and
$\widetilde{E}_{T}(Q^2)$ get positive (negative)
at negative (positive) values of the parameter $t$.

\item In contrary to the isoscalar current case, the values of
$\widetilde{H}_{T}(Q^2)$ are 
positive (negative) for negative (positive) values of $t$.

\item Our final remark is that the LCQSR results on the form factors can be
improved by taking into account the  $\alpha_s$ corrections. 
  
\end{itemize} 

In conclusion, using the most general form of the nucleon interpolating current, we calculate the tensor form factors of nucleon within the LCQSR.  Our results on these form factors are 
compared with the  lattice QCD and chiral soliton
model predictions.\\ \\    
\underline{Note added:}
After completing this work, we become aware of a very recent paper arXiv:1107.4584 [hep-ph] \cite{R10622} in which part of this work is studied.
\section*{Acknowledgment}

We thank P. H\"{a}gler for providing us the lattice QCD data.

\newpage

\begin{figure}
\vskip 2.0 cm
    \includegraphics{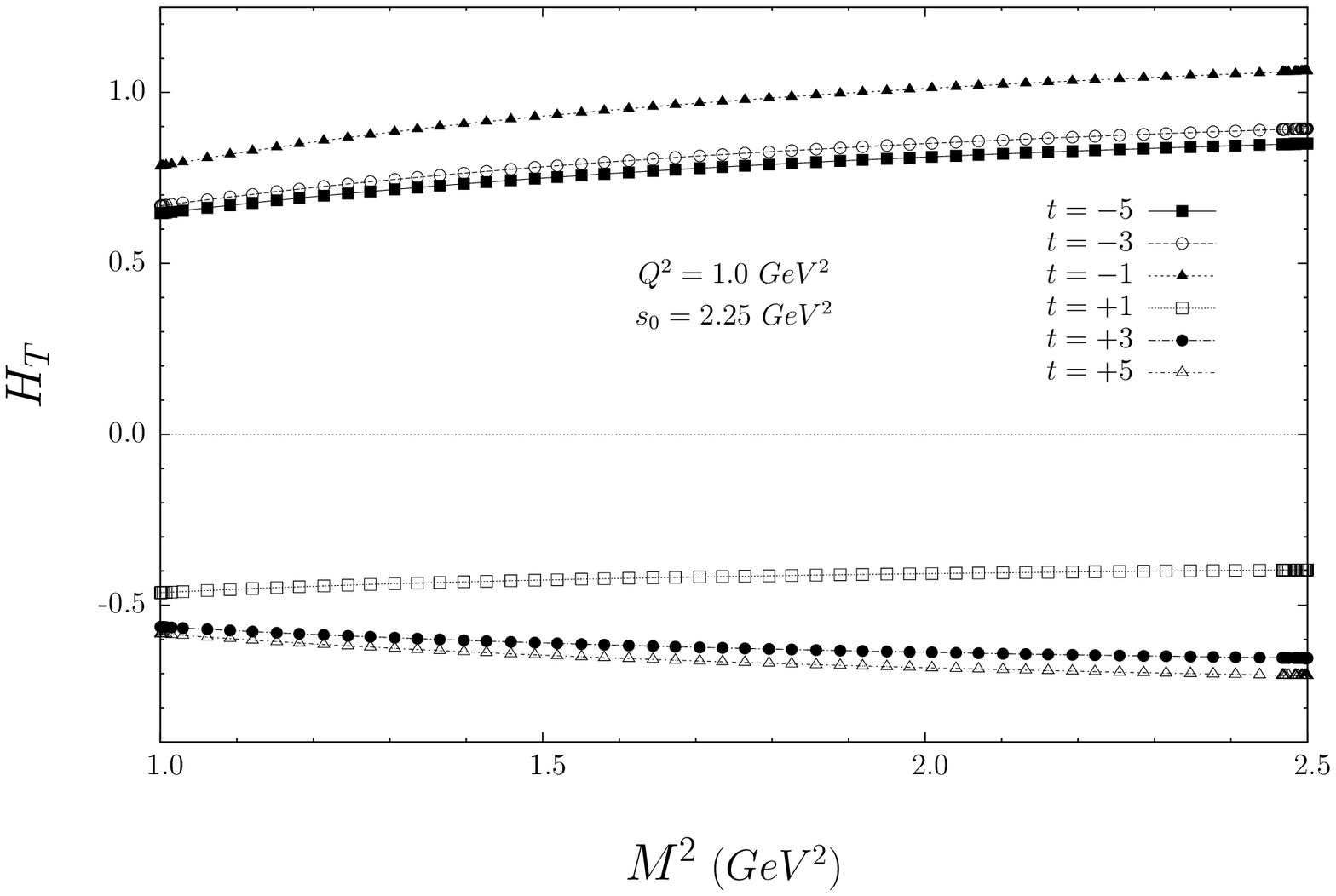}
\vskip 7.5cm
\caption{The dependence of the form factor $H_T$  
of nucleon on $M^2$ at $Q^2=1~GeV^2$ and $s_0=2.25~GeV^2$, 
at six different values of $t$: $t=-5;-3;-1;1;3;5$,
using the first set of DAs for the isoscalar current.}
%\begin{center}
%{\bf Fig. 1--a}
%\end{center}
\end{figure}

\begin{figure}
\vskip 3.0 cm
    \includegraphics{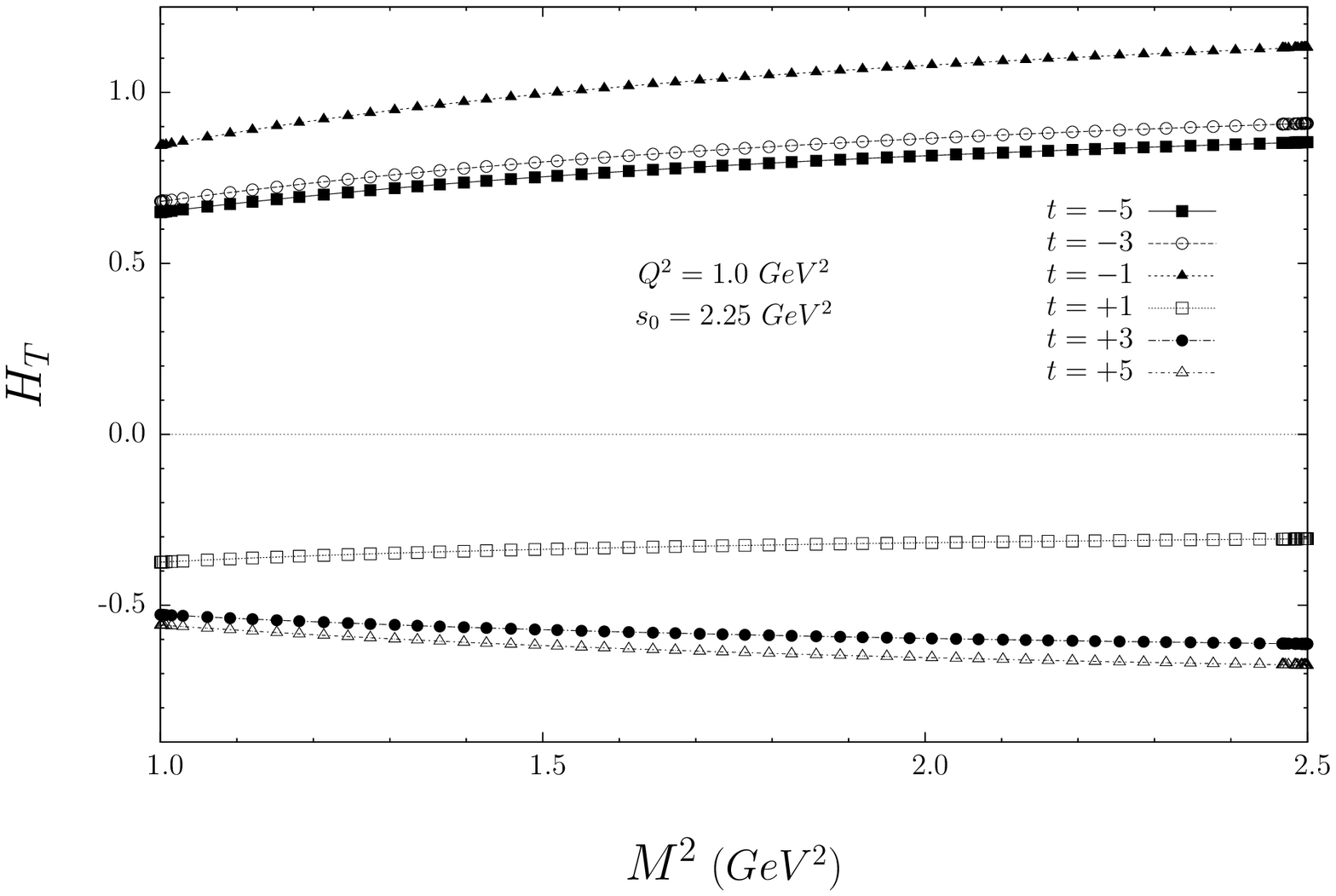}
\vskip 7.5 cm
\caption{The same as in Fig. (1), but at $s_0=2.5~GeV^2$ and 
using the second set of DAs.}
%\begin{center}
%{\bf Fig. 1--b}
%\end{center}
\end{figure}

\begin{figure}
\vskip 2.0 cm
    \includegraphics{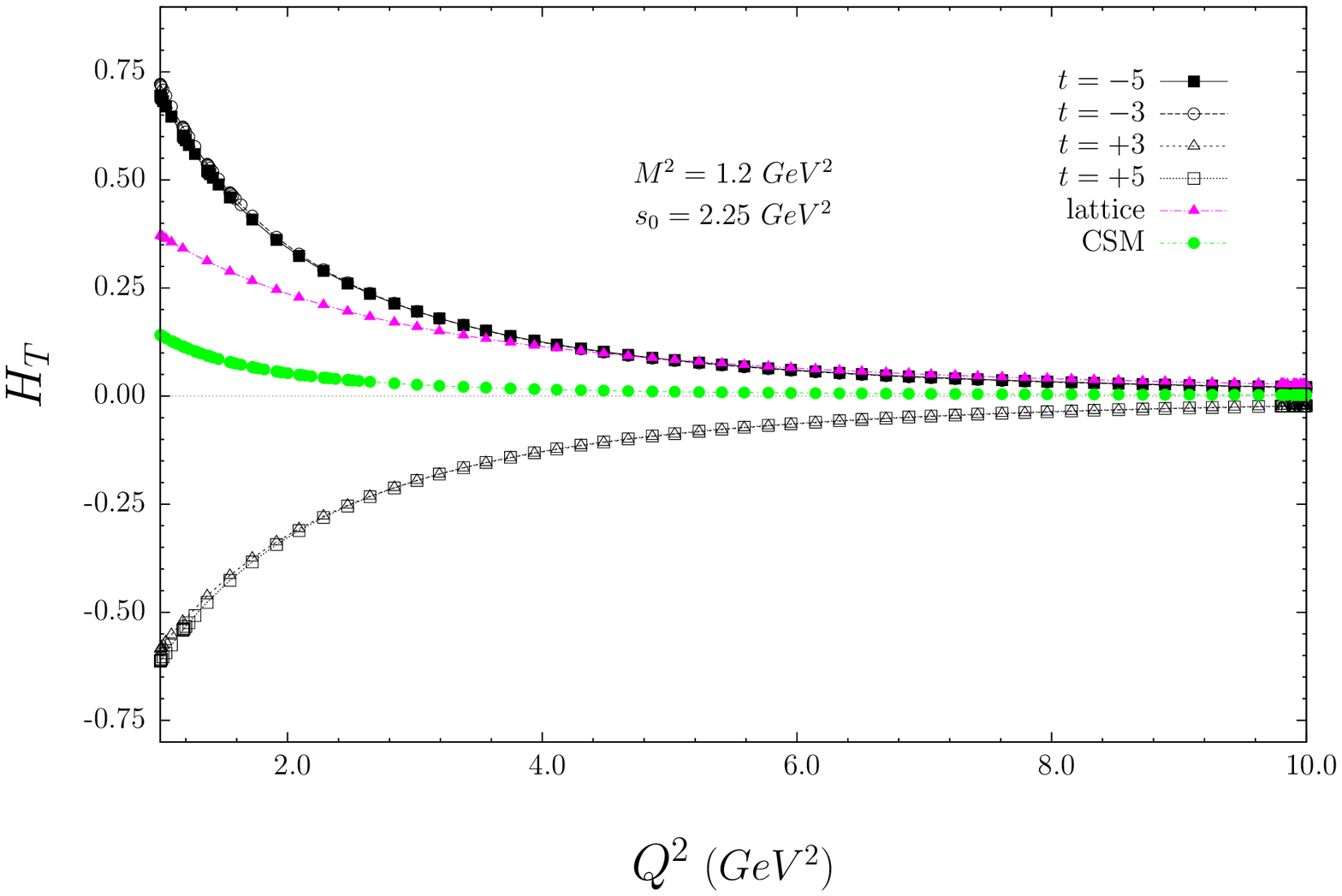}
\vskip 7.5cm
\caption{The dependence of $H_T$ on $Q^2$ at $M^2=1.2~GeV^2$ and
$s_0=2.25~GeV^2$ and four fixed values of $t$: $t=-5;-3;3;5$, for the isoscalar 
current.}
%\begin{center}
%{\bf Fig. 1--a}
%\end{center}
\end{figure}

\begin{figure}
\vskip 3.0 cm
    \includegraphics{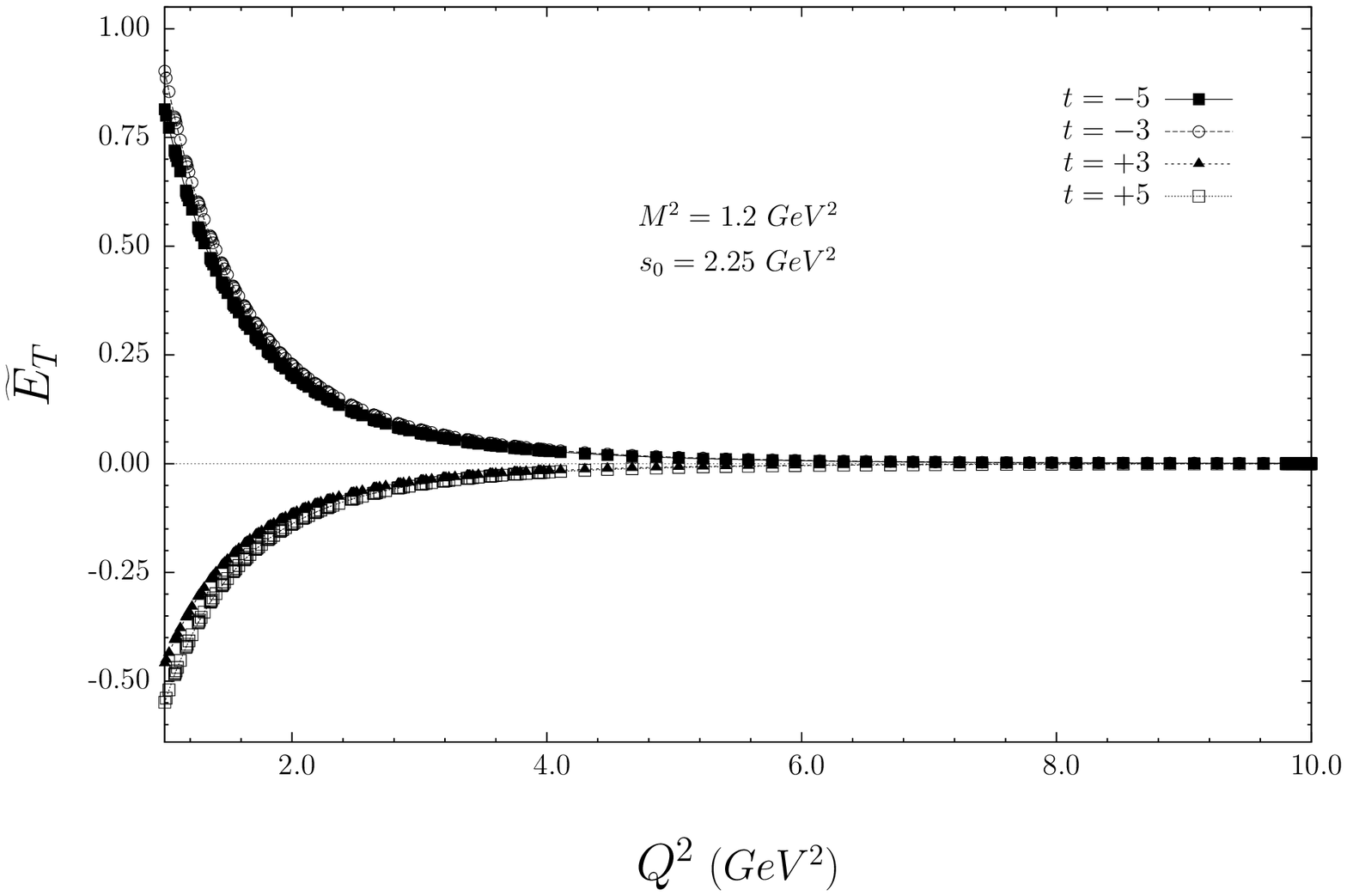}
\vskip 7.5 cm
\caption{The same as in Fig. (3), but for the form factor
$\widetilde{E}_{T}(Q^2)$.}
%\begin{center}
%{\bf Fig. 1--b}
%\end{center}
\end{figure}

\begin{figure}
\vskip 2.0 cm
    \includegraphics{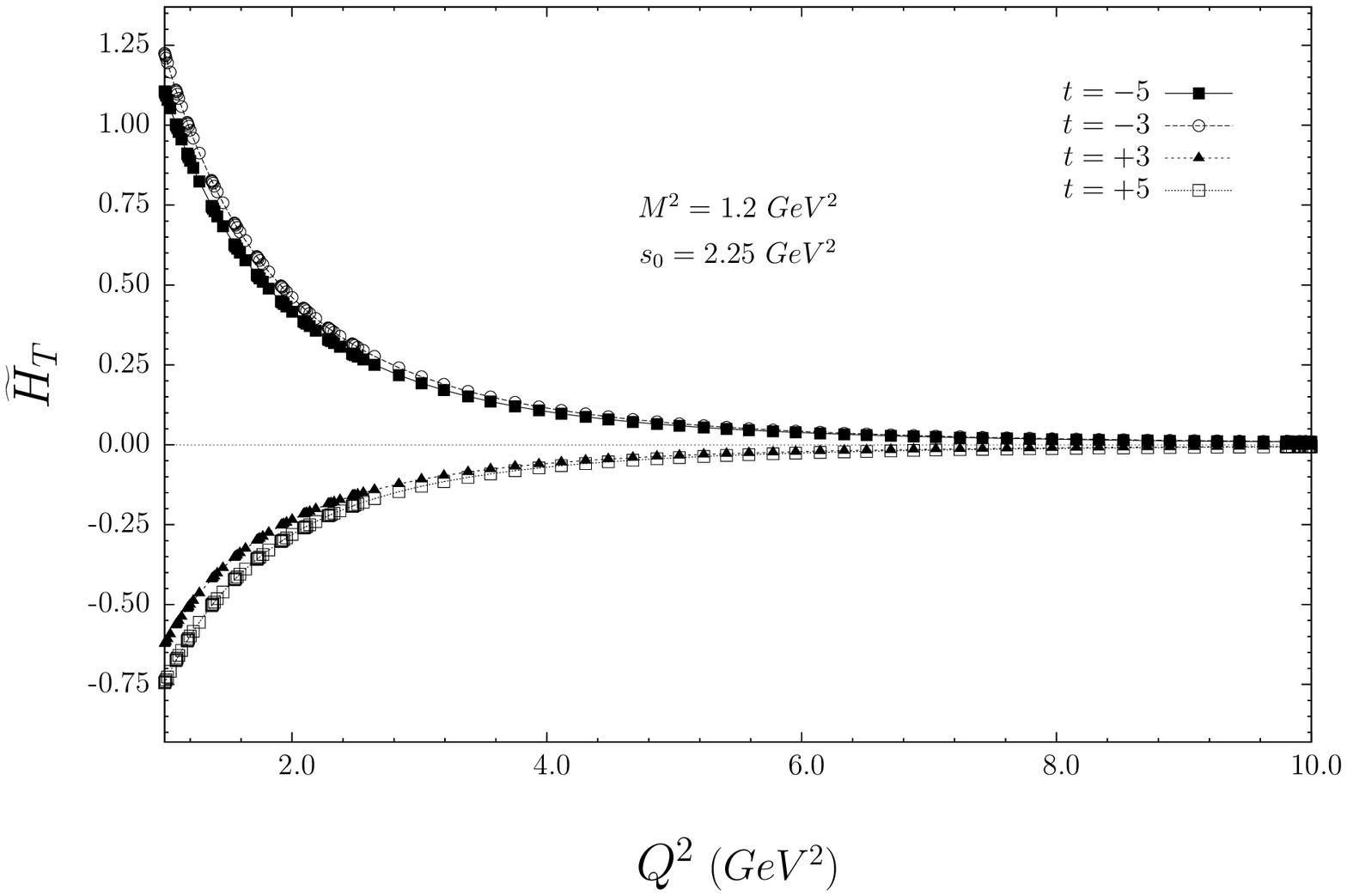}
\vskip 7.5 cm
\caption{The same as in Fig. (3), but for the form factor
$\widetilde{H}_{T}(Q^2)$.}  
%\begin{center}
%{\bf Fig. 1--b}
%\end{center}
\end{figure}                             

\begin{figure}
\vskip 3.0 cm
    \includegraphics{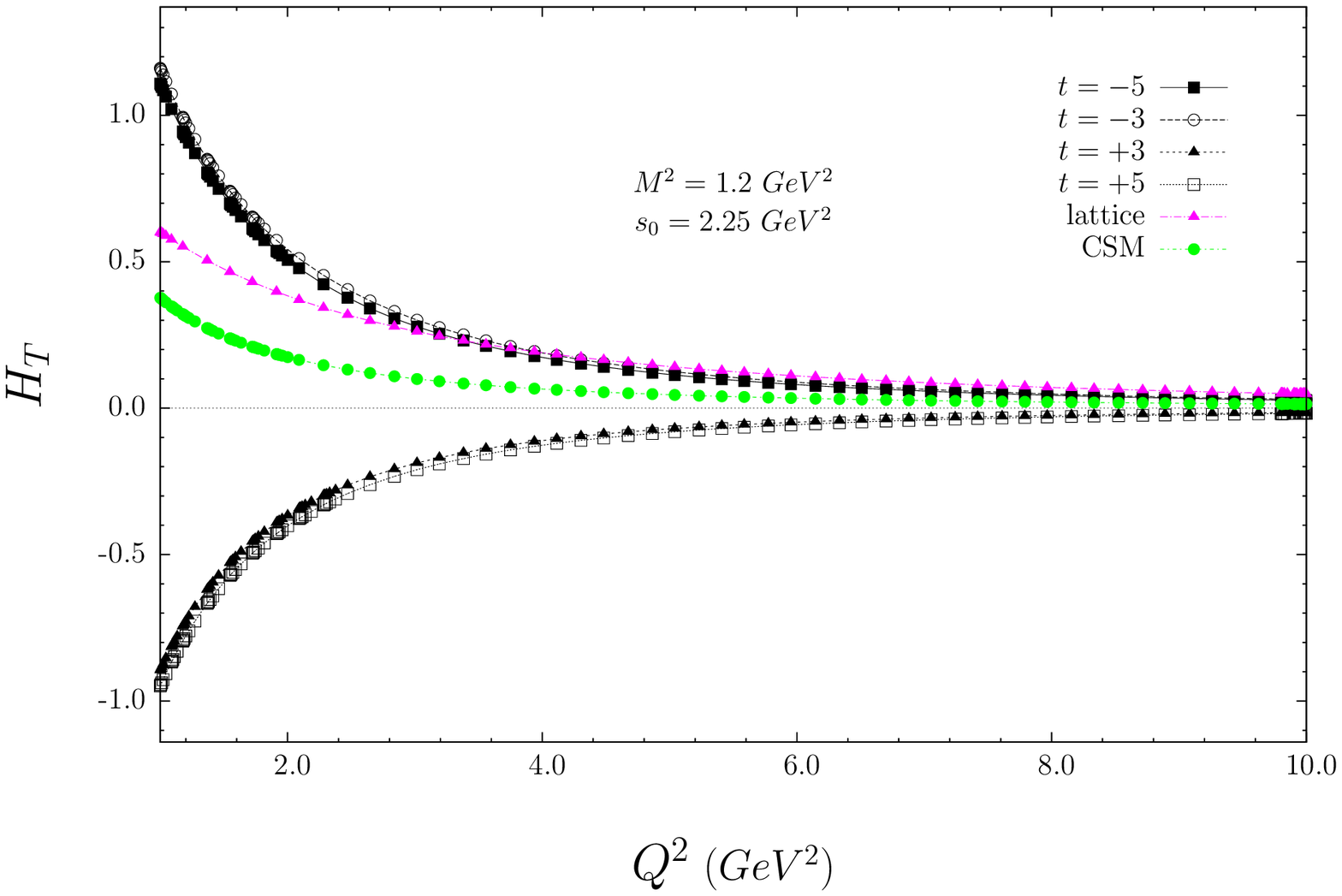}
\vskip 7.5cm
\caption{The same as in Fig. (3), but for the isovector current.}
%\begin{center}
%{\bf Fig. 1--a}
%\end{center}
\end{figure}

\begin{figure}
\vskip 2.0 cm
    \includegraphics{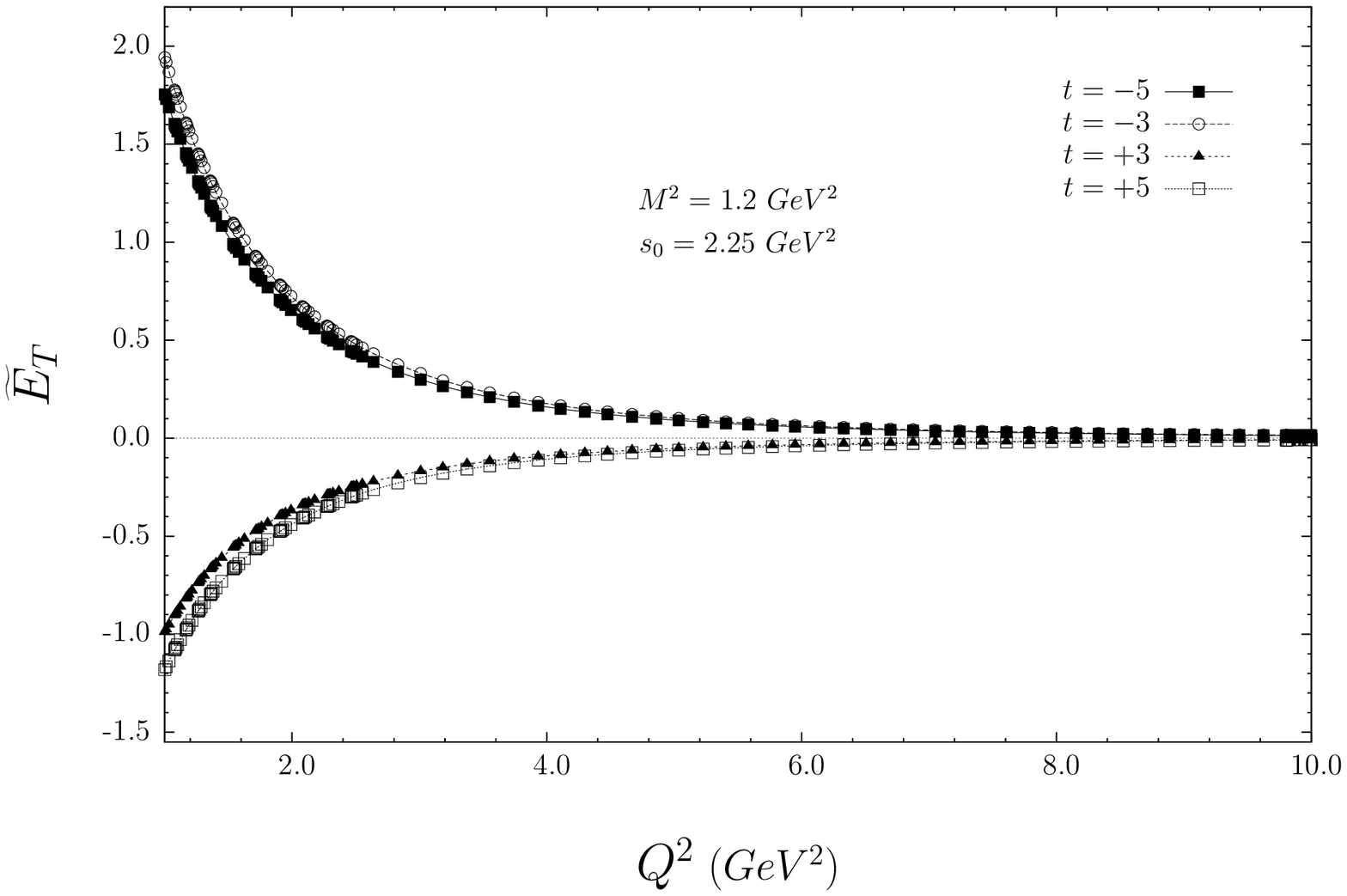}
\vskip 7.5 cm
\caption{The same as in Fig. (4), but for the isovector current.}
%\begin{center}
%{\bf Fig. 1--b}
%\end{center}
\end{figure}

\begin{figure}
\vskip 3.0 cm
    \includegraphics{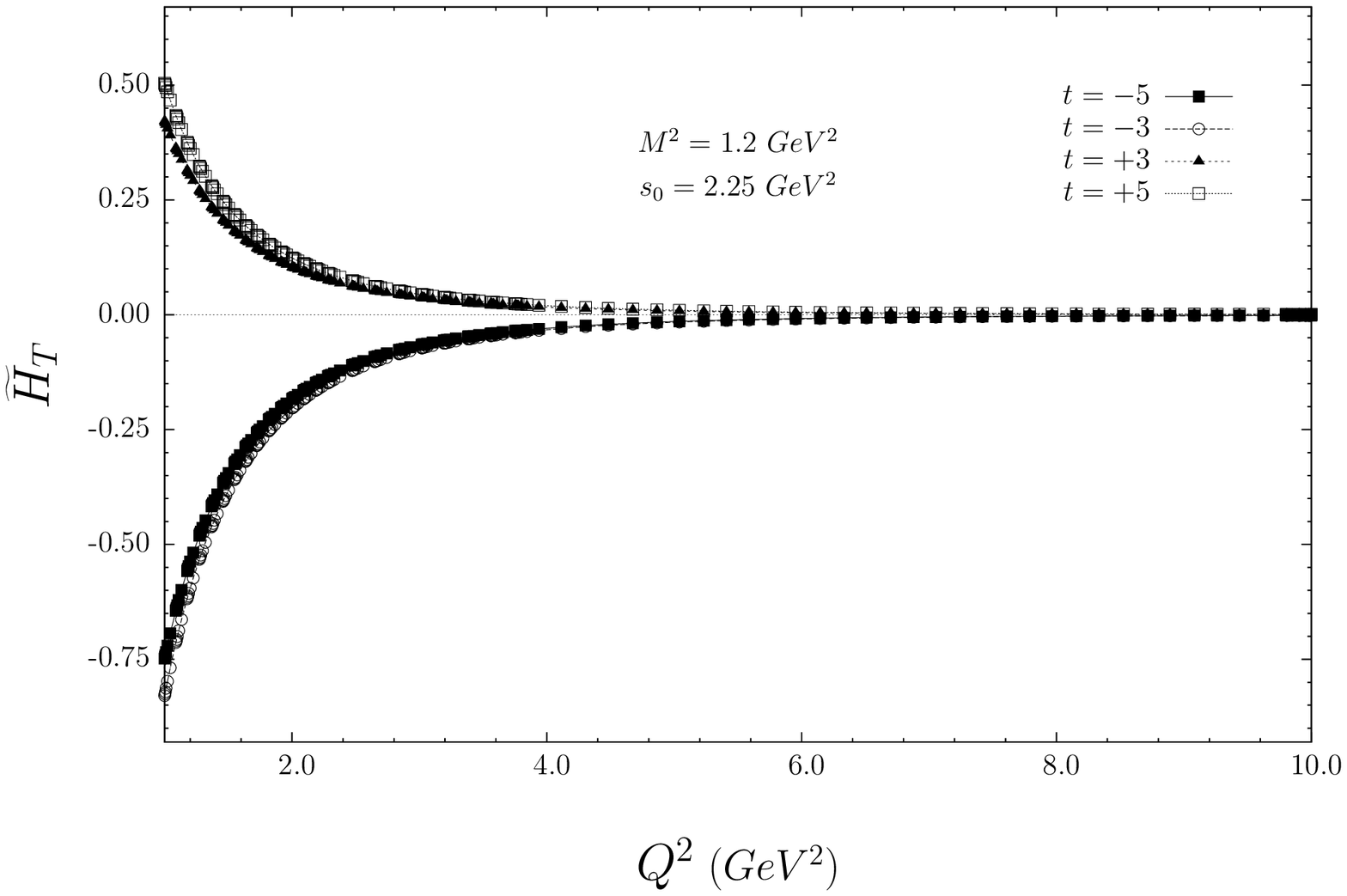}
\vskip 7.5cm
\caption{The same as in Fig. (5), but for the isovector current.}
%\begin{center}
%{\bf Fig. 1--a}
%\end{center}
\end{figure}

\end{document}